\documentclass{article}
\usepackage{graphicx}
\usepackage{amsmath}
\usepackage{amsfonts}
\usepackage{amssymb}

\begin{document}

\author{Antony Valentini\\Augustus College}

\begin{center}
{\LARGE Signal-Locality in Hidden-Variables Theories}

\bigskip

\bigskip\bigskip

\bigskip Antony Valentini\footnote{email: a.valentini@ic.ac.uk}

\bigskip

\bigskip

\textit{Theoretical Physics Group, Blackett Laboratory, Imperial College,
Prince Consort Road, London SW7 2BZ, England.\footnote{Corresponding address.}}

\textit{Center for Gravitational Physics and Geometry, Department of Physics,
The Pennsylvania State University, University Park, PA 16802, USA.}

\textit{Augustus College, 14 Augustus Road, London SW19 6LN,
England.\footnote{Permanent address.}}

\bigskip
\end{center}

\bigskip

\bigskip

\bigskip

We prove that all deterministic hidden-variables theories, that reproduce
quantum theory for a `quantum equilibrium' distribution of hidden variables,
predict the existence of instantaneous signals at the statistical level for
hypothetical `nonequilibrium ensembles'. This signal-locality theorem
generalises yet another property of the pilot-wave theory of de Broglie and
Bohm. The theorem supports the hypothesis that in the remote past the universe
relaxed to a state of statistical equilibrium (at the hidden-variable level)
in which nonlocality happens to be masked by quantum noise.

\bigskip\bigskip

\bigskip

\bigskip


\bigskip

\bigskip

\bigskip

\bigskip

\bigskip

\bigskip

\bigskip

\bigskip

\bigskip

\bigskip

\bigskip

\bigskip

\bigskip

\bigskip

\bigskip

\bigskip

\bigskip

\bigskip

\bigskip

\bigskip

\bigskip

\bigskip

\bigskip

\bigskip

\bigskip

\bigskip

\bigskip

\section{Introduction}

Bell's theorem shows that any reasonable deterministic hidden-variables theory
behind quantum mechanics has to be nonlocal [1].\footnote{It is assumed that
there is no common cause between the hidden variables and the measurement
settings, and no backwards causation (so that the hidden variables are
unaffected by the future outcomes). Bell's original paper addressed only the
deterministic case. The later generalisations to stochastic theories are of no
concern here.} Specifically, for pairs of spin-1/2 particles in the singlet
state, the outcomes of spin measurements at one wing must depend
instantaneously on the axis of measurement at the other, distant wing. In this
paper we show that there are instantaneous signals at the statistical level
for hypothetical ensembles whose distribution differs from that of quantum
theory. This generalises yet another property of de Broglie-Bohm pilot-wave
theory to all deterministic hidden-variables theories.

Historically, Bell's theorem was inspired by the pilot-wave theory of de
Broglie and Bohm [2--10]. Bell asked [11] if all hidden-variables theories
have to be nonlocal like pilot-wave theory; he subsequently proved that they
do [1]. A further property of pilot-wave theory -- `contextuality' -- was
proved to be universal (that is, was proved to be a property of all
hidden-variables theories) by Kochen and Specker [12].\footnote{Even if the
original paper erroneously claimed to prove the nonexistence of hidden
variables. In pilot-wave theory, of course, quantum `measurements' do not
usually reveal the value of some pre-existing attribute of the system, and the
Kochen-Specker theorem shows that all hidden-variables theories must share
this feature.} It is natural to ask if there are any other features of
pilot-wave theory that can be generalised. Here we prove that, indeed,
pilot-wave theory has yet another universal feature: the `signal-locality theorem'.

In pilot-wave theory a system with wavefunction $\psi(x,t)$ has a definite
configuration $x(t)$ at all times, with velocity $\dot{x}(t)=j(x,t)/|\psi
(x,t)|^{2}$ where $j$ is the quantum probability current. To recover quantum
theory one assumes that an ensemble of systems with wavefunction $\psi_{0}(x)$
begins with a distribution of configurations $\rho_{0}(x)=|\psi_{0}(x)|^{2}$
at $t=0$ (guaranteeing $\rho(x,t)=|\psi(x,t)|^{2}$ for all $t$). The Born
probability distribution is assumed as an initial condition. In principle,
however, the theory allows one to consider arbitrary initial distributions
$\rho_{0}(x)\neq|\psi_{0}(x)|^{2}$ which violate quantum theory [5, 10, 13,
14]. (The `quantum equilibrium' distribution $\rho=|\psi|^{2}$ is analogous to
thermal equilibrium in classical mechanics, and may be accounted for by an
\textit{H}-theorem [13, 15].) For two entangled particles at $A$ and $B$ with
wavefunction $\psi(x_{A},x_{B},t)$, operations performed at $B$ (such as
switching on an external potential) have an instantaneous effect on the motion
of the individual particle at $A$. However, for a quantum equilibrium ensemble
$\rho(x_{A},x_{B},t)=|\psi(x_{A},x_{B},t)|^{2}$, operations at $B$ have no
statistical effect at $A$: entanglement cannot be used for signalling at a
distance. But this masking of nonlocality by statistical noise is peculiar to
equilibrium: for an ensemble $\rho_{0}(x_{A},x_{B})\neq|\psi_{0}(x_{A}%
,x_{B})|^{2}$ at $t=0$, changing the Hamiltonian at $B$ generally induces an
instantaneous change in the marginal distribution at $A$ [16].\footnote{The
signal may vanish for some special $\rho_{0}\neq|\psi_{0}|^{2}$, but not in general.}

This is the signal-locality theorem of pilot-wave theory: in general, there
are instantaneous signals at the statistical level if and only if the ensemble
is in quantum nonequilibrium $\rho_{0}\neq|\psi_{0}|^{2}$ [16]. We wish to
show that the same is true in any deterministic hidden-variables theory.

\section{Bell Nonlocality}

Consider two spin-1/2 particles lying on the $y$-axis at $A$ and $B$ and
separated by a large distance. For the singlet state $\left|  \Psi
\right\rangle =\left(  \left|  z+,z-\right\rangle -\left|  z-,z+\right\rangle
\right)  /\surd2$, spin measurements along the $z$-axis at each wing always
yield opposite results. But we are of course free to measure spin components
along arbitrary axes at each wing. For simplicity we take the measurement axes
to lie in the $x-z$ plane; their orientations may then be specified by the
(positive or negative) angles $\theta_{A},$ $\theta_{B}$ made with the
$z$-axis. In units of $\hslash/2$, the possible values of outcomes of spin
measurements along $\theta_{A},$ $\theta_{B}$ at $A,$ $B$ -- that is, the
possible values of the quantum `observables' $\hat{\sigma}_{A},$ $\hat{\sigma
}_{B}$ at $A,$ $B$ -- are $\pm1$. Quantum theory predicts that for an ensemble
of such pairs, the outcomes at $A$ and $B$ are correlated: $\left\langle
\Psi\right|  \hat{\sigma}_{A}\hat{\sigma}_{B}\left|  \Psi\right\rangle
=-\cos(\theta_{A}-\theta_{B})$.

Let us now assume the existence of hidden variables $\lambda$ that determine
the outcomes $\sigma_{A},$ $\sigma_{B}=\pm1$ along $\theta_{A},$ $\theta_{B}$.
And let us assume there exists a `quantum equilibrium ensemble' of $\lambda$
-- that is, a distribution $\rho_{eq}(\lambda)$ that reproduces the quantum
statistics (where $\int d\lambda\ \rho_{eq}(\lambda)=1$). Each value of
$\lambda$ determines a pair of outcomes $\sigma_{A},$ $\sigma_{B}$ (for given
$\theta_{A},$ $\theta_{B})$; for an ensemble of similar experiments -- in
which the values of $\lambda$ generally differ from one run to the next -- one
obtains a distribution of $\sigma_{A},$ $\sigma_{B}$, which is assumed to
agree with quantum theory. In particular, the expectation value%
\[
\overline{\sigma_{A}\sigma_{B}}=\int d\lambda\ \rho_{eq}(\lambda)\sigma
_{A}(\theta_{A},\theta_{B},\lambda)\sigma_{B}(\theta_{A},\theta_{B},\lambda)
\]
must reproduce the quantum result $\left\langle \hat{\sigma}_{A}\hat{\sigma
}_{B}\right\rangle =-\cos(\theta_{A}-\theta_{B})$. Bell showed that this is
possible only if one has nonlocal equations%
\[
\sigma_{A}=\sigma_{A}(\theta_{A},\theta_{B},\lambda),\;\;\;\;\sigma_{B}%
=\sigma_{B}(\theta_{A},\theta_{B},\lambda)
\]
in which the outcomes depend on the distant angular settings
[1].\footnote{Here $\lambda$ are the \textit{initial} values of the hidden
variables, for example just after the source has produced the singlet pair.
Their later values may be affected by changes in $\theta_{A},$ $\theta_{B}$,
and writing $\sigma_{A}=\sigma_{A}(\theta_{A},\theta_{B},\lambda),$
$\sigma_{B}=\sigma_{B}(\theta_{A},\theta_{B},\lambda)$ (where $\lambda$ are
initial values) allows for this. See ref. [4], chapter 8.}

In principle, the nonlocality might be just `one-way', with only one of
$\sigma_{A},$ $\sigma_{B}$ depending on the distant setting. For instance, one
might have $\sigma_{A}=\sigma_{A}(\theta_{A},\theta_{B},\lambda)$ but
$\sigma_{B}=\sigma_{B}(\theta_{B},\lambda)$, with nonlocality from $B$ to $A$
but not from $A$ to $B$.

\section{General Signal-Locality Theorem}

Now given a distribution $\rho_{eq}(\lambda)$, one can always contemplate --
purely theoretically -- a `nonequilibrium' distribution $\rho(\lambda)\neq
\rho_{eq}(\lambda)$, even if one cannot prepare such a distribution in
practice. For example, given an ensemble of values of $\lambda$ with
distribution $\rho_{eq}(\lambda)$, mathematically one could pick a subensemble
such that $\rho(\lambda)\neq\rho_{eq}(\lambda)$.

The theorem to be proved is then the following: in general, there are
instantaneous signals at the statistical level if and only if the ensemble is
in quantum nonequilibrium $\rho(\lambda)\neq\rho_{eq}(\lambda)$.

\textit{Proof}: Assume first that $\sigma_{A}$ has some dependence on the
distant setting $\theta_{B}$. (Bell's theorem requires some nonlocal
dependence in at least one direction.)

Now consider an ensemble of experiments with fixed settings $\theta_{A}%
,\theta_{B}$ and an equilibrium distribution $\rho_{eq}(\lambda)$ of hidden
variables $\lambda$. In each experiment, a particular value of $\lambda$
determines an outcome $\sigma_{A}=\sigma_{A}(\theta_{A},\theta_{B},\lambda)$
at $A$. Some values of $\lambda$ yield $\sigma_{A}=+1$, some yield $\sigma
_{A}=-1$. What happens if the setting $\theta_{B}$ at $B$ is changed to
$\theta_{B}^{\prime}$?

The set $S=\left\{  \lambda\right\}  $ of possible values of $\lambda$ may be
partitioned in two ways:%
\[
S_{A+}=\left\{  \lambda|\sigma_{A}(\theta_{A},\theta_{B},\lambda)=+1\right\}
,\;\;\ S_{A-}=\left\{  \lambda|\sigma_{A}(\theta_{A},\theta_{B},\lambda
)=-1\right\}
\]
where $S=S_{A+}\cup S_{A-}$, $S_{A+}\cap S_{A-}=\emptyset$, and%
\[
S_{A+}^{\prime}=\left\{  \lambda|\sigma_{A}(\theta_{A},\theta_{B}^{\prime
},\lambda)=+1\right\}  ,\;\;\ S_{A-}^{\prime}=\left\{  \lambda|\sigma
_{A}(\theta_{A},\theta_{B}^{\prime},\lambda)=-1\right\}
\]
where $S=S_{A+}^{\prime}\cup S_{A-}^{\prime}$, $S_{A+}^{\prime}\cap
S_{A-}^{\prime}=\emptyset$. (There could exist a pathological subset of $S$
that gives neither outcome $\sigma_{A}=\pm1$, but this must have measure zero
with respect to the equilibrium measure $\rho_{eq}(\lambda)$, and so may be
ignored.) It cannot be the case that $S_{A+}=S_{A+}^{\prime}$ and
$S_{A-}=S_{A-}^{\prime}$ for arbitrary $\theta_{B}^{\prime}$, for otherwise
the outcomes at $A$ would not depend at all on the distant setting at $B$.
Thus in general%
\[
T_{A}(+,-)\equiv S_{A+}\cap S_{A-}^{\prime}\neq\emptyset,\;\ \;\;\;T_{A}%
(-,+)\equiv S_{A-}\cap S_{A+}^{\prime}\neq\emptyset
\]
In other words: under a shift $\theta_{B}\rightarrow\theta_{B}^{\prime}$ in
the setting at $B$, some values of $\lambda$ that would have yielded the
outcome $\sigma_{A}=+1$ at $A$ now yield $\sigma_{A}=-1$; and some $\lambda$
that would have yielded $\sigma_{A}=-1$ now yield $\sigma_{A}=+1$.

Of the equilibrium ensemble with distribution $\rho_{eq}(\lambda)$, a fraction%
\[
\nu_{A}^{eq}(+,-)=\int\nolimits_{T_{A}(+,-)}d\lambda\ \rho_{eq}(\lambda)
\]
make the nonlocal `transition' $\sigma_{A}=+1\rightarrow\sigma_{A}=-1$ under
the distant shift $\theta_{B}\rightarrow\theta_{B}^{\prime}$. Similarly, a
fraction%
\[
\nu_{A}^{eq}(-,+)=\int\nolimits_{T_{A}(-,+)}d\lambda\ \rho_{eq}(\lambda)
\]
make the `transition' $\sigma_{A}=-1\rightarrow\sigma_{A}=+1$ under
$\theta_{B}\rightarrow\theta_{B}^{\prime}$.

Now with the initial setting $\theta_{A},\theta_{B}$, quantum theory tells us
that one half of the equilibrium ensemble of values of $\lambda$ yield
$\sigma_{A}=+1$ and the other half yield $\sigma_{A}=-1$. (That is, the
equilibrium measures of $S_{A+}$ and $S_{A-}$ are both $1/2$.) With the new
setting $\theta_{A},\theta_{B}^{\prime}$, quantum theory again tells us that
one half yield $\sigma_{A}=+1$ and the other half yield $\sigma_{A}=-1$ (the
equilibrium measures of $S_{A+}^{\prime}$ and $S_{A-}^{\prime}$ again being
$1/2$). The 1:1 ratio of outcomes $\sigma_{A}=\pm1$ is preserved under the
shift $\theta_{B}\rightarrow\theta_{B}^{\prime}$, from which we deduce the
condition of `detailed balancing'%
\[
\nu_{A}^{eq}(+,-)=\nu_{A}^{eq}(-,+)
\]
The fraction of the equilibrium ensemble that makes the transition $\sigma
_{A}=+1\rightarrow\sigma_{A}=-1$ must equal the fraction that makes the
reverse transition $\sigma_{A}=-1\rightarrow\sigma_{A}=+1$.

But for an arbitrary \textit{non}equilibrium ensemble with distribution
$\rho(\lambda)\neq\rho_{eq}(\lambda)$, the `transition sets' $T_{A}(+,-)$ and
$T_{A}(-,+)$ will generally have different measures%
\[
\int\nolimits_{T_{A}(+,-)}d\lambda\ \rho(\lambda)\neq\int\nolimits_{T_{A}%
(-,+)}d\lambda\ \rho(\lambda)
\]
and the nonequilibrium transition fractions will generally be unequal,%
\[
\nu_{A}(+,-)\neq\nu_{A}(-,+)
\]

The crucial point here is that $T_{A}(+,-)$ and $T_{A}(-,+)$ are fixed by the
underlying deterministic theory, and are therefore independent of
$\rho(\lambda)$.

Thus, if with the initial setting $\theta_{A},\theta_{B}$ we would have
obtained a certain nonequilibrium ratio of outcomes $\sigma_{A}=\pm1$ at $A$,
with the new setting $\theta_{A},\theta_{B}^{\prime}$ we will in general
obtain a \textit{different} ratio at $A$. Under a shift $\theta_{B}%
\rightarrow\theta_{B}^{\prime}$, the number of systems that change from
$\sigma_{A}=+1$ to $\sigma_{A}=-1$ is unequal to the number that change from
$\sigma_{A}=-1$ to $\sigma_{A}=+1$, causing an imbalance that changes the
outcome ratios at $A$. In other words, in general the statistical distribution
of outcomes at $A$ is altered by the distant shift $\theta_{B}\rightarrow
\theta_{B}^{\prime}$, and there is a statistical signal from $B$ to
$A$.\footnote{The signal vanishes for special $\rho(\lambda)\neq\rho
_{eq}(\lambda)$ that happen to have equal measures for the transition sets,
but not in general.}

Similarly, if $\sigma_{B}$ depends on the distant setting $\theta_{A}$, one
may define non-zero transition sets $T_{B}(+,-)$ and $T_{B}(-,+)$ `from $A$ to
$B$'; and in nonequilibrium there will generally be statistical signals from
$A$ to $B$.

In the special case of `one-way' nonlocality, only one of the pairs
$T_{A}(+,-)$, $T_{A}(-,+)$ or $T_{B}(+,-)$, $T_{B}(-,+)$ has non-zero measure,
and nonequilibrium signalling occurs in one direction only.

\section{Remarks}

The possibility of nonlocal signalling from $B$ to $A$ (or from $A$ to $B$)
depends on the existence of finite transition sets $T_{A}(+,-)$, $T_{A}(-,+)$
(or $T_{B}(+,-)$, $T_{B}(-,+)$). The signal vanishes in equilibrium
$\rho(\lambda)=\rho_{eq}(\lambda)$; while if $\rho(\lambda)$ is concentrated
on just one of $T_{A}(+,-)$, $T_{A}(-,+)$ (or on just one of $T_{B}(+,-)$,
$T_{B}(-,+)$), then all the outcomes are changed by the distant shift. Thus
the size of the signal -- measured by the fraction of outcomes that change at
a distance -- can range from 0\% to 100\%.

It is important to know the size of the transition sets, because if they have
very tiny equilibrium measure, then to obtain an appreciable signal the
nonequilibrium distribution $\rho(\lambda)\neq\rho_{eq}(\lambda)$ would have
to be very far from equilibrium -- that is, concentrated on a very tiny (with
respect to the equilibrium measure) set. Bell's theorem guarantees that at
least one of the pairs $T_{A}(+,-)$, $T_{A}(-,+)$ or $T_{B}(+,-)$,
$T_{B}(-,+)$ has non-zero equilibrium measure (for otherwise we would have a
local theory); but it tells us nothing about the size of these sets: we know
only that, by a detailed-balancing argument, $T_{A}(+,-)$ and $T_{A}(-,+)$
must have equal equilibrium measure, as must $T_{B}(+,-)$ and $T_{B}(-,+)$. In
this sense, Bell's theorem tells us there must be some nonlocality hidden
behind the equilibrium distribution, but not how much.

Given a specific hidden-variables theory, the transition sets may be
determined and their equilibrium measures explicitly calculated. This has been
done in detail for the case of pilot-wave theory [17]. It is found that the
transition sets are generically rather large: for example, in a case where the
measurement procedures are identical at the two wings -- that is, use
identical equipment and coupling -- it is found that with initial settings
$\theta_{A}=\theta_{B}=0$, under a shift $\theta_{B}\rightarrow\theta
_{B}^{\prime}=\pi/2$ at $B$ the equilibrium fraction of outcomes that change
at $A$ is equal to 1/4 (with 1/8 changing from $+1$ to $-1$ and 1/8 from $-1$
to $+1$) [17]. In this case, even a mild disequilibrium $\rho(\lambda)\neq
\rho_{eq}(\lambda)$ entails a significant signal.

But is it possible to derive a general, theory-independent lower bound on the
equilibrium measure of the transition sets?

The problem may be formulated more explicitly as follows. The quantity%
\[
\alpha\equiv\nu_{A}^{eq}(+,-)+\nu_{A}^{eq}(-,+)
\]
(the sum of the equilibrium measures of $T_{A}(+,-)$ and $T_{A}(-,+)$) is the
fraction of the equilibrium ensemble for which the outcomes at $A$ are changed
under $\theta_{B}\rightarrow\theta_{B}^{\prime}$ (irrespective of whether they
change from $+1$ to $-1$ or vice versa, the fractions doing each being
$\alpha/2$). There is a `degree of nonlocality from $B$ to $A$', quantified by
$\alpha=\alpha(\theta_{A},\theta_{B},\theta_{B}^{\prime})$. Similarly, one may
define a `degree of nonlocality from $A$ to $B$', quantified by the fraction
$\beta=\beta(\theta_{A},\theta_{B},\theta_{A}^{\prime})$ of outcomes at $B$
that change in response to a shift $\theta_{A}\rightarrow\theta_{A}^{\prime}$
at $A$.

It is convenient to rewrite $\alpha$ and $\beta$ in a different form. The
quantity $\frac{1}{2}\left|  \sigma_{A}(\theta_{A},\theta_{B}^{\prime}%
,\lambda)-\sigma_{A}(\theta_{A},\theta_{B},\lambda)\right|  $ equals $1$ if
the outcome $\sigma_{A}$ changes under $\theta_{B}\rightarrow\theta
_{B}^{\prime}$, and vanishes otherwise. Since $\rho_{eq}(\lambda)\,d\lambda$
is by definition the fraction of the equilibrium ensemble for which $\lambda$
lies in the interval $(\lambda,\lambda+d\lambda)$, the fraction $\alpha$ for
which $\sigma_{A}$ changes may be expressed as%
\[
\alpha=\frac{1}{2}\int d\lambda\ \rho_{eq}(\lambda)\left|  \sigma_{A}%
(\theta_{A},\theta_{B}^{\prime},\lambda)-\sigma_{A}(\theta_{A},\theta
_{B},\lambda)\right|
\]
Similarly,%
\[
\beta=\frac{1}{2}\int d\lambda\ \rho_{eq}(\lambda)\left|  \sigma_{B}%
(\theta_{A}^{\prime},\theta_{B},\lambda)-\sigma_{B}(\theta_{A},\theta
_{B},\lambda)\right|
\]

Now it is trivial to prove that the `total degree of nonlocality'
$\alpha(\theta_{A},\theta_{B},\theta_{B}^{\prime})+\beta(\theta_{A},\theta
_{B},\theta_{A}^{\prime})$ cannot vanish for all settings $\theta_{A}%
,\theta_{B},\theta_{A}^{\prime},\theta_{B}^{\prime}$. For $\alpha$ and $\beta$
are both non-negative; so if $\alpha+\beta=0$ then $\alpha=\beta=0$; and if
this were true for all settings, $\sigma_{A}(\theta_{A},\theta_{B},\lambda)$
would have no dependence on $\theta_{B}$ and $\sigma_{B}(\theta_{A},\theta
_{B},\lambda)$ would have no dependence on $\theta_{A}$ (apart from a possible
set of $\lambda$ of equilibrium measure zero), in contradiction with Bell's
theorem. But is it possible to derive some sort of lower bound on
$\alpha+\beta$?

Positive lower bounds on $\alpha+\beta$, and on $\alpha$ or $\beta$ alone,
have been derived assuming certain symmetries; and it has been checked in
detail that these bounds are satisfied by pilot-wave theory [17]. But a
general lower bound has not yet been obtained.

It should be possible to derive -- without any extra assumptions -- a general
lower bound on the \textit{average} total degree of nonlocality $\overline
{\alpha+\beta}$, obtained by averaging $\alpha+\beta$ over all possible
initial and final settings $\theta_{A},\theta_{B},\theta_{A}^{\prime}%
,\theta_{B}^{\prime}$.\footnote{Alternatively, one might look at the maximum
value $(\alpha+\beta)_{\text{max}}$.} Again, Bell's theorem trivially implies
that $\overline{\alpha+\beta}>0$, but something more is needed to obtain a
positive lower bound.

Note that $\alpha$ and $\beta$ have a natural interpretation in terms of
nonlocal information flow at the hidden-variable level. For an equilibrium
ensemble of pairs, shifting the angle at $B$ alters a fraction $\alpha$ of the
outcomes at $A$ (from $+1$ to $-1$ or vice versa). Thus $\alpha$ may be
interpreted as the \textit{average number of bits of information} per singlet
pair transmitted nonlocally (in equilibrium) from $B$ to $A$, and similarly
for $\beta$ from $A$ to $B$ [17]. Of course, in equilibrium this information
flow is not visible at the statistical level, because as many outcomes flip
from $+1$ to $-1$ as from $-1$ to $+1$. But each individual change in outcome
does represent, at the hidden-variable level, the transmission of one bit of
`subquantum information'.

Finally, we remark that since the signal-locality theorem generalises yet
another property of pilot-wave theory (in addition to nonlocality and
contextuality), one wonders if there remain still further properties of
pilot-wave theory that are actually universal properties of hidden-variables
theories generally.

\section{Conclusion and Hypothesis}

We have proved a general `signal-locality theorem': in any deterministic
hidden-variables theory that reproduces quantum statistics for some
`equilibrium' distribution $\rho_{eq}(\lambda)$ of hidden variables $\lambda$,
a generic `nonequilibrium' distribution $\rho(\lambda)\neq\rho_{eq}(\lambda)$
would give rise to instantaneous signals at the statistical level (as occurs
in pilot-wave theory).

Bell's theorem tells us that if hidden variables exist then so do
instantaneous influences. But there is no consensus on what to conclude from
this. Similarly, one must distinguish between the signal-locality theorem
proved above and what this author proposes to conclude from it.

It seems mysterious that nonlocality should be hidden by an all-pervading
quantum noise. We have shown that any deviation from that noise would make
nonlocality visible. It is as if there is a conspiracy in the laws of physics
that prevents us from using nonlocality for signalling. But another way of
looking at the matter is to suppose that our universe is in a state of
statistical equilibrium at the hidden-variable level, a special state in which
nonlocality happens to be hidden. The physics we see is not fundamental; it is
merely a phenomenological description of an equilibrium state [16].

This view is arguably supported by quantum field theory in curved spacetime,
where there is no clear distinction between quantum and thermal fluctuations
[18]. On this basis it has been argued that quantum and thermal fluctuations
are really the same thing [19]. This suggests that quantum theory is indeed
just the theory of an equilibrium state, analogous to thermal equilibrium.

On this view it is natural to make the hypothesis that the universe began in a
state of quantum nonequilibrium $\rho(\lambda)\neq\rho_{eq}(\lambda)$, where
nonlocal signalling was possible, the relaxation $\rho(\lambda)\rightarrow
\rho_{eq}(\lambda)$ taking place during the great violence of the big bang [5,
10, 13, 14, 16]. In effect, a hidden-variables analogue of the classical
thermal heat death has actually occurred in our universe. (In the classical
heat death, thermal energy cannot be used to do work; in the `quantum heat
death', nonlocality cannot be used for signalling.)

This hypothesis could have observable consequences. The nonlocal effects of
disequilibrium may have played a role in homogenising the universe at early
times [10]. In cosmological inflationary theories, early corrections to
quantum fluctuations would change the spectrum of primordial density
perturbations imprinted on the cosmic microwave background [10, 20]. And
particles that decoupled at sufficiently early times could still be in quantum
nonequilibrium today: thus, exotic particles left over from the very early
universe might violate quantum mechanics [10, 13, 14, 20].

\textbf{Acknowledgements.} For helpful comments and discussions I am grateful
to Guido Bacciagaluppi, Jossi Berkovitz, Lucien Hardy, Lee Smolin and
Sebastiano Sonego, to audiences at the Universities of Maryland, Notre Dame
and Utrecht, and to participants at the NATO workshop `Modality, Probability,
and Bell's Theorems', Cracow, August 19--23, 2001. This work was supported by
the Jesse Phillips Foundation.

\bigskip\bigskip

\begin{center}
\textbf{REFERENCES}
\end{center}

\bigskip

[1] J.S. Bell, Physics 1 (1965) 195.

[2] L. de Broglie, in: Electrons et photons (Gauthier-Villars, Paris, 1928).
[English translation: G. Bacciagaluppi and A.Valentini, Electrons and photons:
the proceedings of the fifth solvay congress (Cambridge University Press, forthcoming).]

[3] D. Bohm, Phys. Rev. 85 (1952) 166, 180.

[4] J.S. Bell, Speakable and unspeakable in quantum mechanics (Cambridge
University Press, 1987).

[5] A. Valentini, On the pilot-wave theory of classical, quantum and
subquantum physics, PhD Thesis (1992), International School for Advanced
Studies, Trieste.

[6] P. Holland, The quantum theory of motion (Cambridge University Press, 1993).

[7] D. Bohm and B.J. Hiley, The undivided universe (Routledge, 1993).

[8] J.T. Cushing, Quantum mechanics (Chicago, 1994).

[9] Bohmian mechanics and quantum theory, eds. J.T. Cushing \textit{et al.}
(Kluwer, 1996).

[10] A. Valentini,\textit{ }Pilot-wave theory of physics and cosmology
(Cambridge University Press, forthcoming).

[11] J.S. Bell, Rev. Mod. Phys. 38 (1966) 447.

[12] S. Kochen and E. Specker, J. Math. Mech. 17 (1967) 59.

[13] A. Valentini, in: Chance in physics, eds. J. Bricmont \textit{et al}.
(Springer, 2001).

[14] A. Valentini, in ref. [9].

[15] A. Valentini, Phys. Lett. A 156 (1991) 5.

[16] A. Valentini, Phys. Lett. A 158 (1991) 1.

[17] A. Valentini, in: Modality, probability, and Bell's theorems, eds. T.
Placek and J. Butterfield (Kluwer, 2002) [quant-ph/0112151].

[18] D.W. Sciama, P. Candelas, and D. Deutsch, Adv. Phys. 30 (1981) 327.

[19] L. Smolin, Class. Quantum Grav. 3 (1986) 347.

[20] A. Valentini, Int. J. Mod. Phys. A, forthcoming.
\end{document}